# Towards Designing a Biometric Measure for Enhancing ATM Security in Nigeria E-Banking System

Ibidapo, O. Akinyemi, Zaccheous O. Omogbadegun, and Olufemi M. Oyelami

*Abstract*—Security measures at banks can play a critical, contributory role in preventing attacks on customers. These measures are of paramount importance when considering vulnerabilities and causation in civil litigation. Banks must meet certain standards in order to ensure a safe and secure banking environment for their customers. This paper focuses on vulnerabilities and the increasing wave of criminal activities occurring at Automated Teller Machines (ATMs) where quick cash is the prime target for criminals rather than at banks themselves. A biometric measure as a means of enhancing the security has emerged from the discourse.

*Keywords*—Security, ATM, Biometric, Crime.

## I. INTRODUCTION

Automated teller machine is a mechanical device that has its roots embedded in the accounts and records of a banking institution [1]. It is a machine that allows the banks customers carry out banking transactions like, deposits, transfers, balance enquiries, and withdrawal. Notwithstanding, we lived in a world where people no longer want to encounter long queues for any reason, they don't not want to wait on queue for too long a time before they are attended to and this has led to the increasing services being rendered by banks to further improve the convenience of banking through the means of electronic banking. On this note the advent of ATM is imperative, although with its own flaws.

Crime at ATM's has become a nationwide issue that faces not only customers, but also bank operators [2]. Security measures at banks can play a critical, contributory role in preventing attacks on customers. These measures are of paramount importance when considering vulnerabilities and causation in civil litigation and banks must meet certain standards in order to ensure a safe and secure banking environment for their customers.

Basically, the ATM scam involves thieves putting a thin, clear, rigid plastic sleeve into the ATM card slot. When you insert your card, the machine can't read the strip, so it keeps asking you to re-enter your PIN number. Meanwhile, someone behind you watches as you tap in your number. Eventually you give up, thinking the machine has swallowed your card and you walk away. The thieves then remove the plastic sleeve complete with card, and empty your account. The way to avoid this is to run your finger along the card slot before you put your card in. The sleeve has a couple of tiny prongs that the thieves need to get the sleeve out of the slot, and you'll be able to feel them. The primary focus of this work is on developing a biometric strategy (Fingerprint) to enhance the security features of the ATM for effective banking transaction. The rest of the paper arranged thus: section 2 presents the methodology employed for the work, section 3 examines related work, section 4 presents the system design and the general concept of the implementation, and section 5 concludes the work

## II. METHODOLOGY

The security feature for enhancing the ATM was designed using the client/server approach. There will be a link between the customer's identification information, customer's accounts and records in the bank (server). The network is designed to support a large number of users and uses dedicated server to accomplish this. The reason for choosing Client/Server model for this application is because it provides adequate security for the resources required for a critical application such as Banking. Similarly, a descriptive conceptual approach which includes Unified Modeling language (UML) tools such as Use case models, class diagrams etc is adapted. Microsoft Access 2003 as a database software is employed to create database to store cardholder's information. The work is implemented using Visual Basic 6.0 software tool, used to design the user interfaces and/or cardholder interaction with the ATM Machine.

## III. ELECTRONIC BANKING

E-banking can be defined as the deployment of banking services and products over electronic and communication networks directly to customers [3]. It is the automated delivery of new and traditional banking products and services directly to customers through electronic, interactive communication channels [4].

These electronic and communication networks include Automated Teller Machines (ATMs), direct dial-up connections, private and public networks, the Internet, televisions, mobile devices and telephones. Among these technologies, the increasing penetration of personal computers, relatively easier access to the Internet and particularly the wider diffusion of mobile phones has drawn the attention of most banks to e-banking. However, the continuing convergence of information, communications and media technologies is also opening up new electronic channels (such as "*pod-banking*") of delivering banking services.

Significant differences exist among banks in terms of their e-banking capabilities. These differences can take two main dimensions. The first is the use of electronic channels and the second is the sophistication of banking services delivered over an electronic channel. Many established banks in developed countries began with ATMs and evolved through Personal Computer-banking, Telephone-banking, Internet-banking, TV-banking, and Mobile-banking. However, this evolution is not visible in recently established banks and in most of the African countries with the exception of South Africa. It appears that e-banking has dawned in Africa with Internet-banking [5].

E-banking systems can vary significantly in their configuration depending on a number of factors. Financial institutions should choose their e-banking system configuration, including outsourcing relationships, based on four factors, therefore strategic objectives for E-Banking; scope, scale, and complexity of equipment, systems, and activities; technology expertise; and security and internal control requirements In terms of e-banking services sophistication, this ranges from one way *information-push* services where customers receive information about the bank, its products and services to *information-download* where customers can download (or ask in case of telephone-banking) account information and forms to *full-transaction* services where customers can perform most banking transactions (such as transfer between accounts, bill payment, third party payment, card and loan applications, etc) electronically (see for example [6], [4]. Some banks do also provide new banking products (such as e-saving) that are only accessible electronically [2]. Some of the key drivers of offering e-banking services include reducing transaction costs, increasing convenience, availability and timeliness of transactions, and improving accessibility for better fund administration [5]. Achieving these objectives tend to contribute strategic benefits in terms of better customer relationship management, increased customer base, and improved market image [2]. 2006).

In Nigeria today, all the financial services industry has been subject to dramatic changes over the past three years, as a result of advances in IT, capitalization, deregulation, and globalization. These changes have reduced margins in traditional banking activities, leading banks to merge with other banks. The forces of consolidation are also having a profound impact on the operation of securities exchanges, as well as the brokerage and asset management industries [7]. The merger programme of the banking institutions has resulted in the consolidation of 51 banking institutions into 10 banking groups. The mergers, which involved the consolidation of 96% of the total assets of the banking institutions was achieved with minimum disruption and dislocation to the system. This has been a major accomplishment by the domestic banking industry. The domestic banking groups are now in a position to reap greater benefits from economies of scale, through greater investment in technology and the more substantive pool of skilled staff. This will allow the banking institutions to make further gains on efficiency and competitiveness.

As a result of this development, virtually all the new 25 banks offers Internet banking facilities while one (GTBank) has introduced mobile banking. The adoption of chip technology now also offers new forms of payment choices and higher security to the public. The introduction of the magnetic stripe ATM cards is indeed an important step forward.

## IV. SYSTEM DESIGN AND IMPLEMENTATION

This research is being carried out for the sole purpose of designing a three factor authentication metrics, that is, the ATM ID number, the PIN number and the Biometric feature (fingerprint). It is expected that the customer should possess an ATM card, to know and remember his/her PIN number and to enroll his/her fingerprint into the fingerprint device/reader adapter into the system. After which the fingerprint database compares the live sample provided by the customer with the template in the database. On confirmation that the information provided is true, that customer is granted access to the ATM system. For the design of this system Unified Modeling language tools (use case models and activity diagram) to represent how the user (bank customer) interacts with the proposed system are employed.

Use cases are scenarios for understanding system requirements. A use-case model can be instrumental in project development, planning, and documentation of system requirements. A use case is an interaction between users and a system; it captures the goal of the users and the responsibility of the system to its users. It describes the uses of the system and shows the courses of events that can be performed as well as defining what happens in a system. In essence, the use case model tries to systematically identify uses of the system and therefore the system's responsibilities. A use case model provides an external view of a system or application; it is directed towards the users or the "actors" of the systems, not its implementers. In the design of the ATM application, the actor of the bank system is the bank customer. The bank customer must be able to deposit certain amount to and withdraw any amount from his or her accounts (provided he/she has up such amount in the account) using the bank application. Figure 1 below show the use case diagram for our system design, where customers can perform transaction by inserting their ATM card and carry out the Approval Process by entering PIN Number and Confirm Fingerprint. After the approval, customer is requested type of transaction (deposit of money or withdrawal of money), and the transaction is carried out accordingly. At the completion of the transaction, the customer exit Application and remove his/her card. A detail description of the system is shown in the activity diagram in figure2.

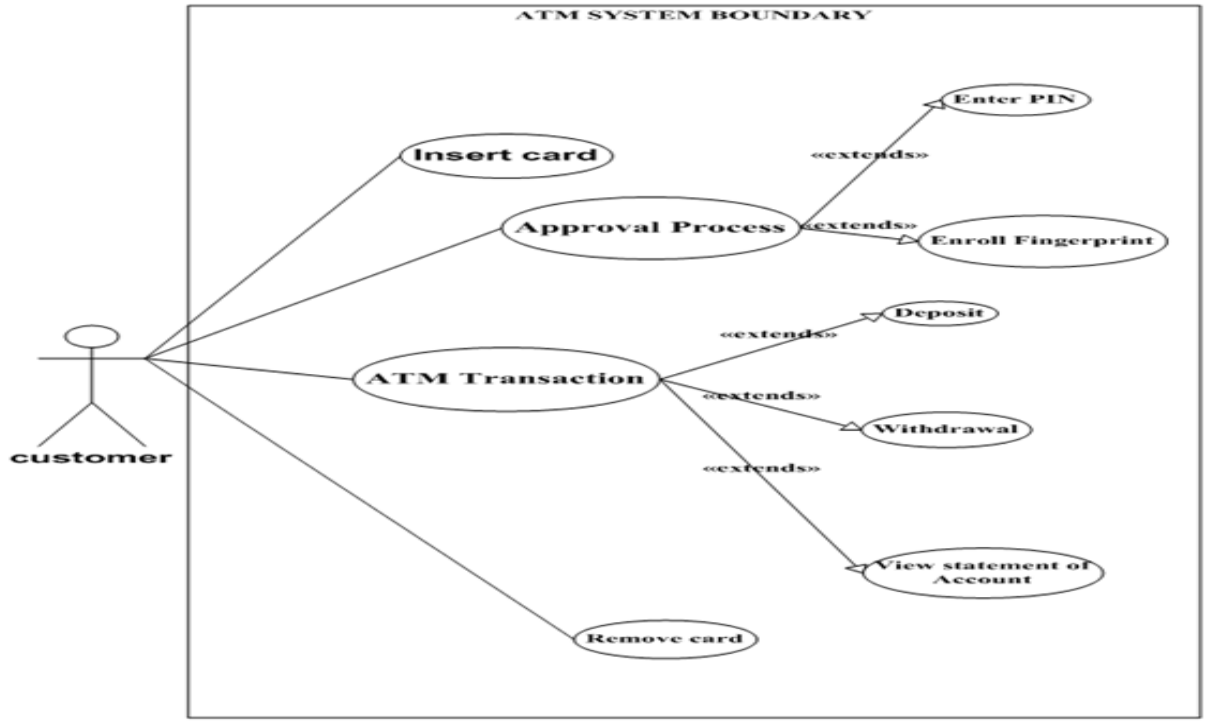

Fig.1: Use Case Diagram for customer's transaction.

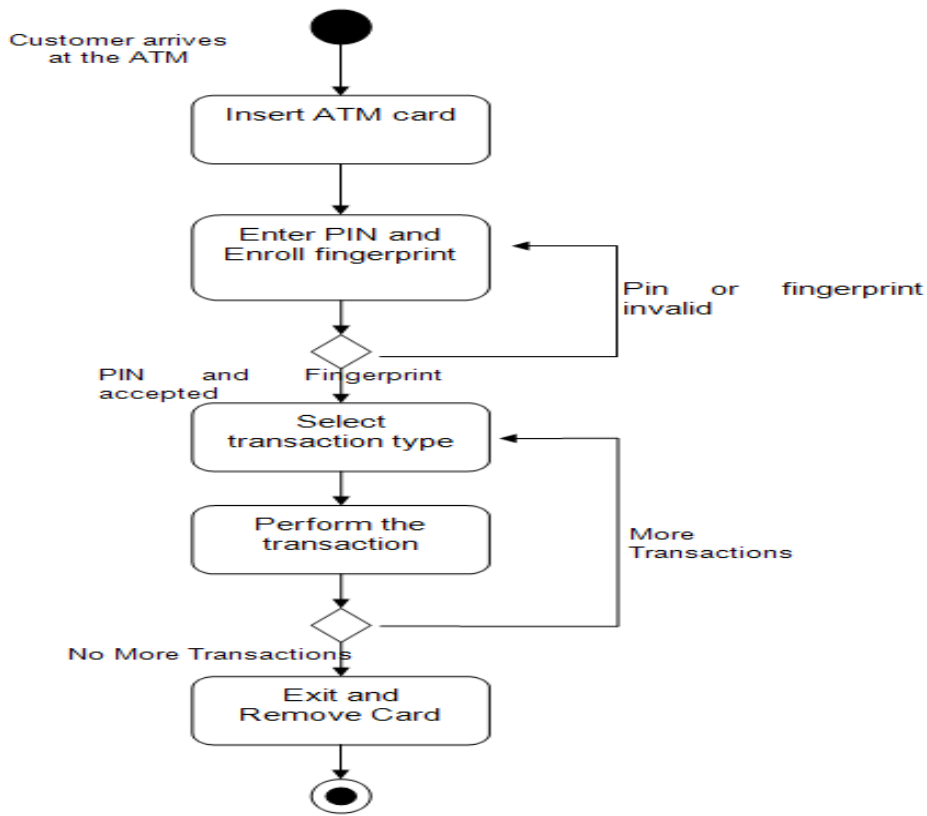

Fig. 2: Activity Diagram for customer's transaction.

*A. User Interface Desig*

A user interface is a friendly means by which users of a system can interact with the system to process inputs and obtain outputs. It is also a means of communication between the human user and the system through the use of input/output devices with supporting software. This particular ATM application is made up of 6 interfaces, which include; Login Interface, Enroll Fingerprint Interface, Transaction Type Selection Interface, Withdrawal Interface, Deposit Interface, and View statement of Account Interface.

*1. User Interface Design*

This interface is the very first interface the bank customer interacts with on the ATM machine. This interface prompts the customer to insert ATM card and proceeds with the entire authentication processes, that is, inputting the ID (or card number) and PIN number (see figure 3). If the user enters an invalid card number or PIN number, a dialogue box appears prompting an invalid PIN or invalid card number and the system returns enter a valid PIN number. A typical description of this is shown in figure 4. After validating the customer's card and PIN number, the customer is directed to the next phase of the authentication process via the authentication dialogue box for inputting the fingerprint.

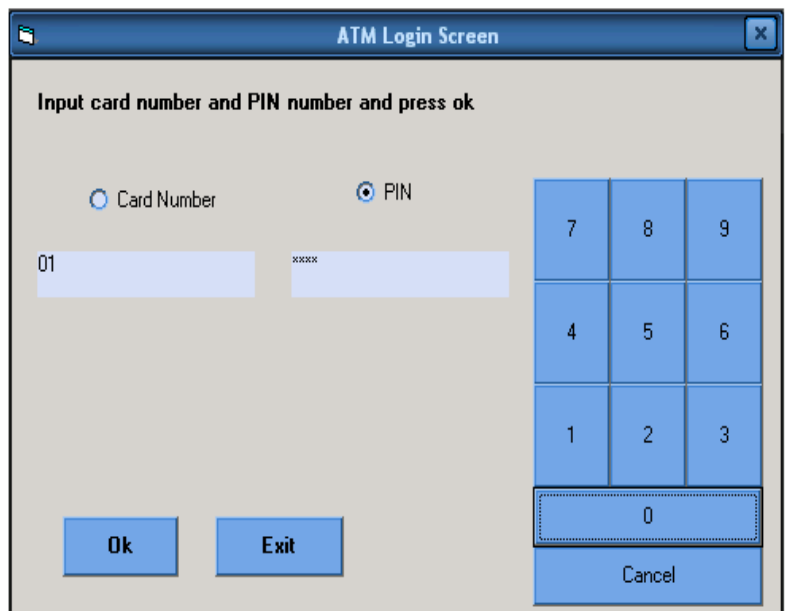

Fig 3: Login Interface

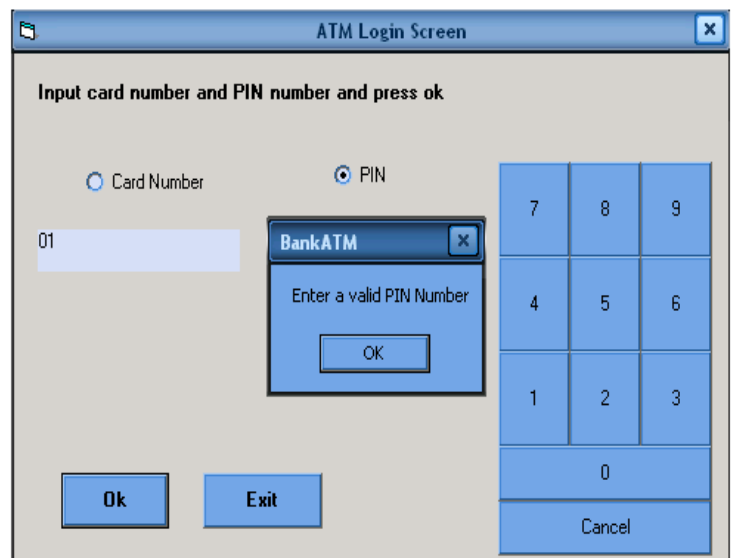

Fig 4: Login Interface (response to invalid card number/PIN)

*2. Fingerprint Interface*

This is the final interface the customer interacts with in the authentication process. It requests from the customer the enrollment of his/her fingerprint to be placed on a Fingerprint reader. The fingerprint reader accepts the fingerprint and seeks to match the live sample with the already enrolled templates in the banks database. If match is confirmed it will finally authenticate customer else it will deny customer access to his/her bank account.

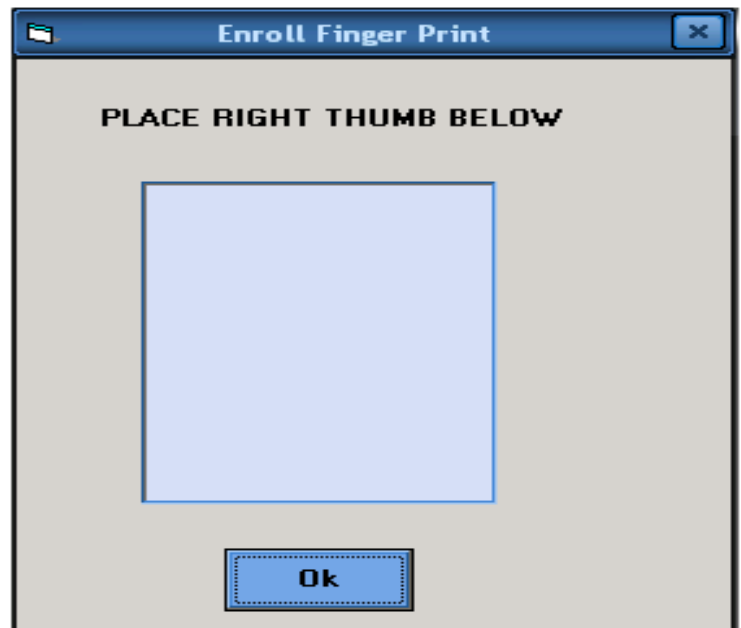

Fig 5: Fingerprint Interface

The fingerprint of an individual is very peculiar to that individual since no two individuals can have the same fingerprint. The fingerprint reader captures the fingerprint features of an individual and search for a match of fingerprint brought up for identification among the stored fingerprints in the database. . The fingerprints stored are kept along side the other ID's (Pin and Card Numbers) and the corresponding biometric templates are kept in the database. When the fingerprint is found correct, the customer is taken to the transaction phase where he/she will choose among the transactions (deposit or withdrawal), otherwise the customer is denied access and the system brings up a dialogue box for which the customer can choose Ok, and as soon as this done the system automatically log off the customer. Figure 5 below depicts this behavior.

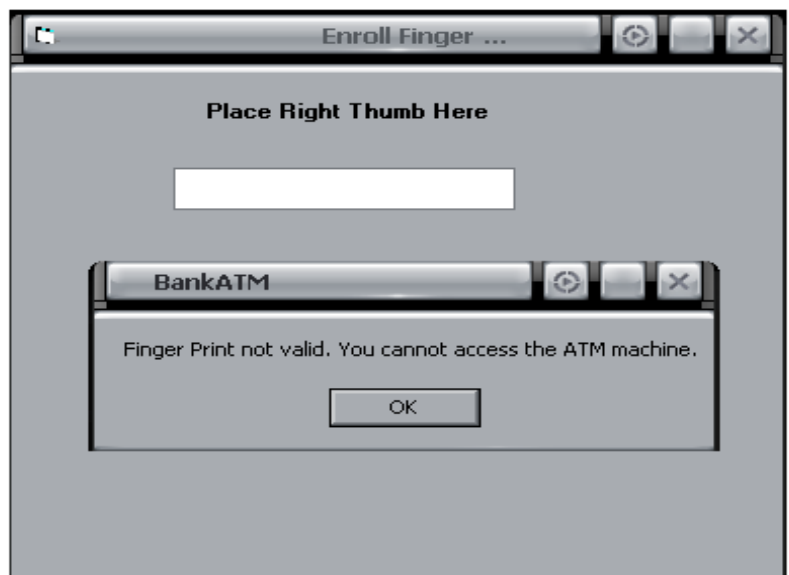

Fig 6  Invalid Fingerprint

*3. Withdrawal Interface*

This interface enables the customer withdraw money from his/her account. It shows the customers current balance by subtracting the amount withdrawn from the previous account balance. After the customer has completed all his/her withdrawals, a dialogue box pops up notifying the customer of his/her successful withdrawal transaction. The interface is shown below.

Fig 6: Withdrawal Interface

I. CONCLUSION

We have been able to develop a fingerprint mechanism as a biometric measure to enhance the security features of the ATM for effective banking transaction for Banks in Nigeria. The prototype of the developed application has been found promising on the account of its sensitivity to the recognition of the customers' finger print as contained in the database. This system when fully deployed will definitely reduce the rate of fraudulent activities on the ATM machines such that only the registered owner of a card access to the bank account.